 \documentclass[final,3p,times,twocolumn]{elsarticle}

\usepackage{graphicx}
\usepackage{soul,color}
\usepackage{amssymb}

\usepackage{lineno}

\usepackage{booktabs}

\journal{arXiv}

\begin{document}

\begin{frontmatter}


\title{Asymmetries in football: The pass-goal paradox}



\author[madrid1,madrid2,madrid3]{D.R. Antequera}
\author[madrid1,madrid2,madrid3]{D. Garrido}
\author[madrid1,madrid2,madrid3]{I. Echegoyen}
\author[laliga]{R. L\'opez del Campo}
\author[laliga]{R. Resta Serra}
\author[madrid1,madrid2,madrid3,madrid4]{J. M. Buld\'u}

\address[madrid1]{Complex Systems Group, Universidad Rey Juan Carlos, Madrid, Spain}
\address[madrid2]{Grupo Interdisciplinar de Sistemas Complejos (GISC), Spain}
\address[madrid3]{Laboratory of Biological Networks, Center for Biomedical Technology, Universidad Polit\'ecnica de Madrid, Madrid, Spain}
\address[laliga]{Mediacoach - LaLiga, Madrid, Spain}
\address[madrid4]{Institute of Unmanned System and Center for OPTical IMagery Analysis and Learning (OPTIMAL), Northwestern Polytechnical University, Xi'an 710072, China}

\begin{abstract}
We investigate the relation between the number of passes made by a football team and the number of goals. We analyze the 380 
matches of a complete season of the Spanish national league {\em ``LaLiga"} (2018/2019). We observe how the number of scored goals is positively correlated with the number of passes made by a team. In this way, teams on the top (bottom) of the ranking at the end of the season make more (less) passes than the rest of the teams. However, we observe a strong asymmetry when the analysis is made depending on the part of the match. Interestingly, fewer passes are made on the second part of a match while, at the same time, more goals are scored. This paradox appears in the majority of teams, and it is independent of the number of passes made. These results confirm that goals in the first part of matches are more ``costly" in terms of passes than those scored on second halves.
\end{abstract}


\end{frontmatter}


\section{Introduction}
\label{S:1}

Year after year, the analysis of actions and patterns occurring in a football match is becoming more complex \cite{buldu2018,ribeiro2020,salmon2020}. 
Technology is the main responsible for the avalanche of new
kind of datasets that analysts and data scientists working in football clubs have to deal with \cite{gudmundsson2017}. 
In this way, every action occurring in the pitch is recorded and categorized, from passes to goals, but also
tackles, shots, fouls, corners, possessions... At the same time, the position of all players (including the referees) 
and the ball is recorded at rates up to 25 frames per second, which allows obtaining
not only the position of players in real-time but also their speeds, accelerations, or total distances covered.

The availability of these datasets has resulted in a diversity of new kind of methodologies and metrics to understand what
is happening on the pitch. New points of view have arisen, such as evaluating the control of the pitch \cite{fernandez2018}, measuring the area covered by the convex hull \cite{arrudamoura2013} or tracking the evolution of the passing networks between players \cite{buldu2018}. Furthermore, new metrics have been defined to quantify the performance of specific actions such as the expected goal (xG)  parameter \cite{rathke2017,spearman2018}, which quantifies the quality of a shot, or the post-shot expected goals (PSxG), defined for evaluating goalkeepers \cite{psxg2020}.

However, despite the increasing complexity of the analysis in football, there are still fascinating conclusions drawn from a closer inspection
of the classical football indicators \cite{mackenzie2013}. For example, Lago-Pe\~nas et al. analyzed the final result of a match when the home (or away) team scored first \cite{lago2016}. They
showed that teams that scored first ended the match scoring around the double of their opponents. Furthermore, home teams scored first around 60\% of the matches.
Another approach is to count the number of passes. In \cite{hughes2005}, authors 
counted the passes made before goals during the 1990 Fifa World Cup finals, showing
that successful teams scored more goals after longer passing sequences. In a more recent study analyzing the 2004 European Championship, Yiannakos and Armatas 
showed the existence of a high percentage of long passes before goals
but, more importantly, they reported a higher percentage of goals in 
the second part ($57.4\%$) than in the first part of the match ($42.6\%$) \cite{yiannakos2006}, a fact also observed in other studies \cite{alberti2013,leite2013}.

Redwood-Brown went one step beyond and investigated the number and accuracy of passes before and after scoring a goal \cite{redwood2008}. Interestingly, he observed that, during the five minutes before a goal, the number of passes was higher than the average. On the contrary, during the five minutes after a goal, the number of
passes was lower. Furthermore, the accuracy of passes was also related to scoring, 
with teams showing a higher percentage of successful passes before scoring a goal and a lower percentage during the following five minutes \cite{redwood2008}.

In this paper, we investigated the relation between the number of passes made by a team and the number of goals. We analyzed the 380 matches
of the 2018/2019 season of the Spanish national football league  {\em ``LaLiga"}. Our analysis focused on two issues, first, we wanted to confirm the results
presented by Redwood-Brown \cite{redwood2008}, which suggested that increasing the number of passes could be related with increasing the probability of scoring a goal. Second, we investigated the differences between the first and second parts of a match, intending to find analogies/discrepancies between them. Our results show that, indeed, there is a relation between the number of passes and scored goals, although the correlation between
both variables (passes and goals) was not as high as we expected. However, we found an interesting paradox when looking at the differences between parts: Despite passes
and goals have a positive correlation between them, second parts have a lower number of passes while, at the same time, the number of goals is higher. In this way, the number of passes required to score a goal is much higher in the first part of a match, making passes of the second parts more efficient. 

\section{Results}

\subsection{More passes, more goals}

\begin{figure}[!t] 
 \centering
 \includegraphics[width=0.49\textwidth]{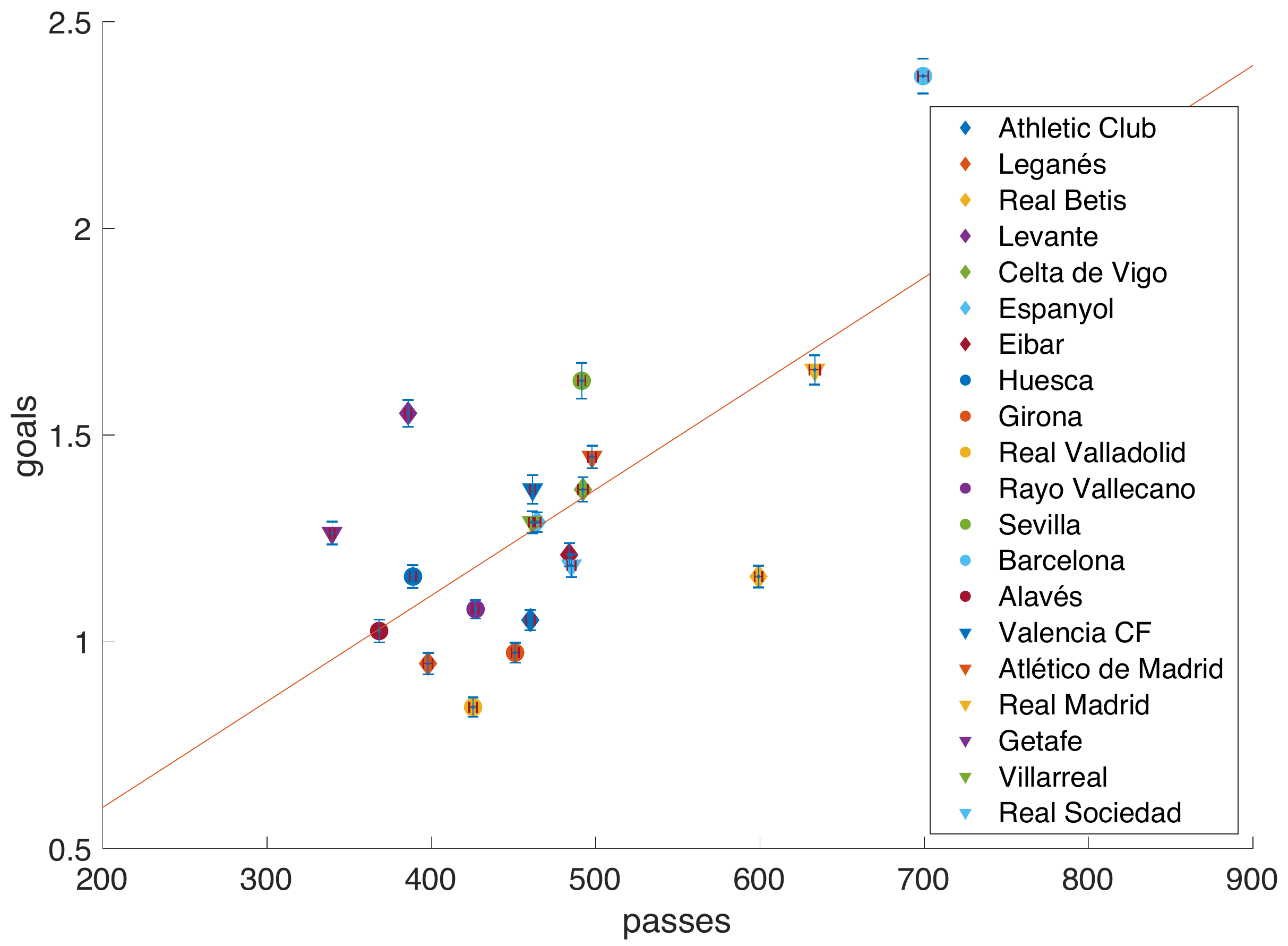}
 \caption{
\textbf{Correlation between the number of completed passes and scored goals.} Each point corresponds to a team and the red solid line is the linear regression
of the points, which has a slope $m=0.026$, an intercept of $b=0.087$ and a correlation coefficient of $r=0.672$.
}\label{f01}
\end{figure}

The datasets we analyzed consisted of the number of passes and goals made by each of the $20$ teams participating at the Spanish national football league ({\em ``LaLiga" Santander}).
Specifically, we have a total of $N=357724$ completed passes and $M=983$ goals.
We also considered the temporal information (minute and second) of both types of events, which allowed us to separate between the first and second halves of the match.
Figure \ref{f01} shows the number of completed passes made by each team vs. the number of scored goals. The solid red line is the linear regression of the data, which had a correlation coefficient of $r=0.6724$. It seems that there is a positive correlation between both variables, although its value is rather low. 
However, this 
result is not conclusive, so let us carry out an alternative analysis to shed more light on the interplay between passes and goals. Table \ref{tab01}
shows the average value of the number of passes grouped in 3 different categories: (1) teams that finished in the top four (T4), 
which qualified for the European Champions league, 
(2) teams in the middle ranking (MR), from position 5th to position 17th and (3) the three teams that were relegated (RE) to the second division. 
We can observe how teams on the top four have the highest average number of passes, followed by the teams in the middle 
of the table and, finally, relegated teams. On the second column Tab.\ref{tab01}, we show
the average number of goals for each group. Comparing both columns, we can observe that the higher the number of 
passes of a group, the higher the number of scored
goals and, furthermore, the higher the position at the final raking. 

\begin{table}[b]
\centering
\begin{tabular}{lcc}
\hline

\noalign{\smallskip}
Group & Passes & Goals  \\
\noalign{\smallskip}
\bottomrule[1.2pt]

\noalign{\smallskip}
Top 4 (T4) & $573 \pm 112$  & $1.71 \pm 0.45$   \\
\noalign{\smallskip}

\noalign{\smallskip}
Middle Raking (MR)& $450 \pm 68$  & $1.21 \pm 0.22$ \\
\noalign{\smallskip}

\noalign{\smallskip}
Relegated Teams (RE) & $422 \pm 31$  & $1.07 \pm 0.09$  \\
\noalign{\smallskip}

\hline
\end{tabular}
\caption{\textbf{Average number of passes and goals.} Teams are divided into three categories: (1) teams at the top 4 (T4), which were qualified for
playing the European Champions League, (2) the 13 teams at the middle of the ranking (MR) and (3) the 3 teams that were relegated (RE) to a lower division. Numbers correspond
to the mean number of passes and goals per match and their corresponding standard deviation.
}\label{tab01}
\end{table}

Are these results statistically significant? To answer this question we considered the variables ``pass" and ``goal" obtained for 
all matches of teams belonging to each group. 
We had to randomly sample $114$ values at each group, since groups have different number of observations and we were limited by the number 
of observations of the smallest group. 
Then, we run a 1-way ANOVA to compare the passes of the three groups and a 1-way Kruskal-Wallis (KW) test to compare their goals. The latter is a 
non-parametric approach to the former, given that the number of goals per match is very low, and thus we cannot expect it to follow a normal distribution. Then, we compared groups in pairs, to check if they have equal means/medians or not. 
Finally, to ensure that the statistical analysis was unbiased, we repeated this process $1000$ times (sampling, general test, pair-wise comparisons), correcting the p-values for multiple comparisons with False Discovery Rate, adjusted for $\alpha = 0.01$ \cite{benjamini2005false}.


\begin{table}[!b]
\centering
\hspace{-0.6cm}
\begin{tabular}{lccc}
\hline

\noalign{\smallskip}
Groups & $\mu_{diff} (\pm \sigma_{diff})$ & $\mu_{p-val} $ & $\%_{sig}$ \\
\noalign{\smallskip}
\bottomrule[1.2pt]

\noalign{\smallskip}
T4-RE & $ 150.46 (\pm 13.18) $ & $9.56e^{-10} $ & $ 100 \%$ \\
\noalign{\smallskip}

\noalign{\smallskip}
T4-MR & $ 121.66 (\pm 16.28) $ & $ 9.39e^{-09}$ & $ 100 \%$ \\
\noalign{\smallskip}

\noalign{\smallskip}
MR-RE & $ 28.79 (\pm 9.52) $ & $0.19 $ & $0 \%$  \\
\noalign{\smallskip}

\hline
\end{tabular}
\caption{\textbf{Statistical differences in passes between groups}. Teams are divided into three categories: (1) teams at the top 4 (T4), which were qualified for playing the European Champions League, (2) the 13 teams at the middle of the ranking (MR) and (3) the 3 teams that were relegated to a lower division (RE). Each row considers a pair of groups (T4-RE, T4-MR and MR-RE), for which we show the average ($\pm $ standard deviation) difference in passes across all sampling iteration (1000 in total; see main text for details), as well as the average p-value. Third column shows the percentage of cases in which the statistical comparison between groups rejected the null hypothesis of equal means. 
}\label{tab02}
\end{table}

Tables \ref{tab02} and \ref{tab03} show the results of the group comparisons in passes and goals, respectively. From left to right: (i) average difference (standard deviation) between groups, (ii) average p-value associated to it, and (iii) percentage of iterations (out of $1000$) in which we can safely state that there are statistical differences between groups. Note that all p-values shown hereafter have been already corrected for multiple comparisons. 
As we can see in Tabs. \ref{tab02}-\ref{tab03}, differences between relegated and middle ranking teams are not statistically significant, no matter the variable used to compare them (goals/passes). On the contrary, top 4 teams are clearly different to the other two groups in terms of passes (100\% of cases in which we find statistically significant differences after correcting for multiple comparisons). Differences are one order of magnitude higher in these cases. Concerning the number 
of goals (Tab. \ref{tab03}.),
differences are not as evident, but some of them fulfill the statistical tests.  
Relegated and top 4 teams show statistical significant differences in a high percentage of cases (79.4\%), with p-values of 0.01. The comparison between the top 4 teams 
and middle ranking teams show statistical significant differences in the 33.5\% of the cases with a p-value of 0.09. Taking all into account, we can conclude that top 4 teams
make more passes and goals than the rest, while relegated teams are those with the lower number of passes and goals. 



\begin{table}[t]
\centering
\begin{tabular}{lccc}
\hline

\noalign{\smallskip}
Groups & $\mu_{diff} (\pm \sigma_{diff})$ & $\mu_{p-val} $ & $\%_{sig}$ \\
\noalign{\smallskip}
\bottomrule[1.2pt]

\noalign{\smallskip}
T4-RE & $ 45.29 (\pm 9.56) $ & $ 0.01 $ & $79.4 \%$ \\
\noalign{\smallskip}

\noalign{\smallskip}
T4-MR & $ 35.29 (\pm 12.81)$ & $ 0.09 $ & $33.5 \%$ \\
\noalign{\smallskip}

\noalign{\smallskip}
MR-RE & $ 10 (\pm 8.7) $ & $ 0.64 $ & $0 \%$  \\
\noalign{\smallskip}

\hline
\end{tabular}
\caption{\textbf{Statistical differences in goals between groups}. Teams are divided into three categories: (1) teams at the top 4 (T4), which were qualified for playing the European Champions League, (2) the 13 teams at the middle of the ranking (MR) and (3) the 3 teams that were relegated to a lower division (R). Each row considers a pair of groups (T4-RE, T4-MR and MR-RE), for which we show the average ($\pm $ standard deviation) difference in passes across all sampling iteration (1000 in total; see main text for details), as well as the average p-value. Third column shows the percentage of cases in which the statistical comparison between groups rejected the null hypothesis of equal means. 
}\label{tab03}
\end{table}

\begin{figure}[!b] 
 \centering
 \includegraphics[width=0.48\textwidth]{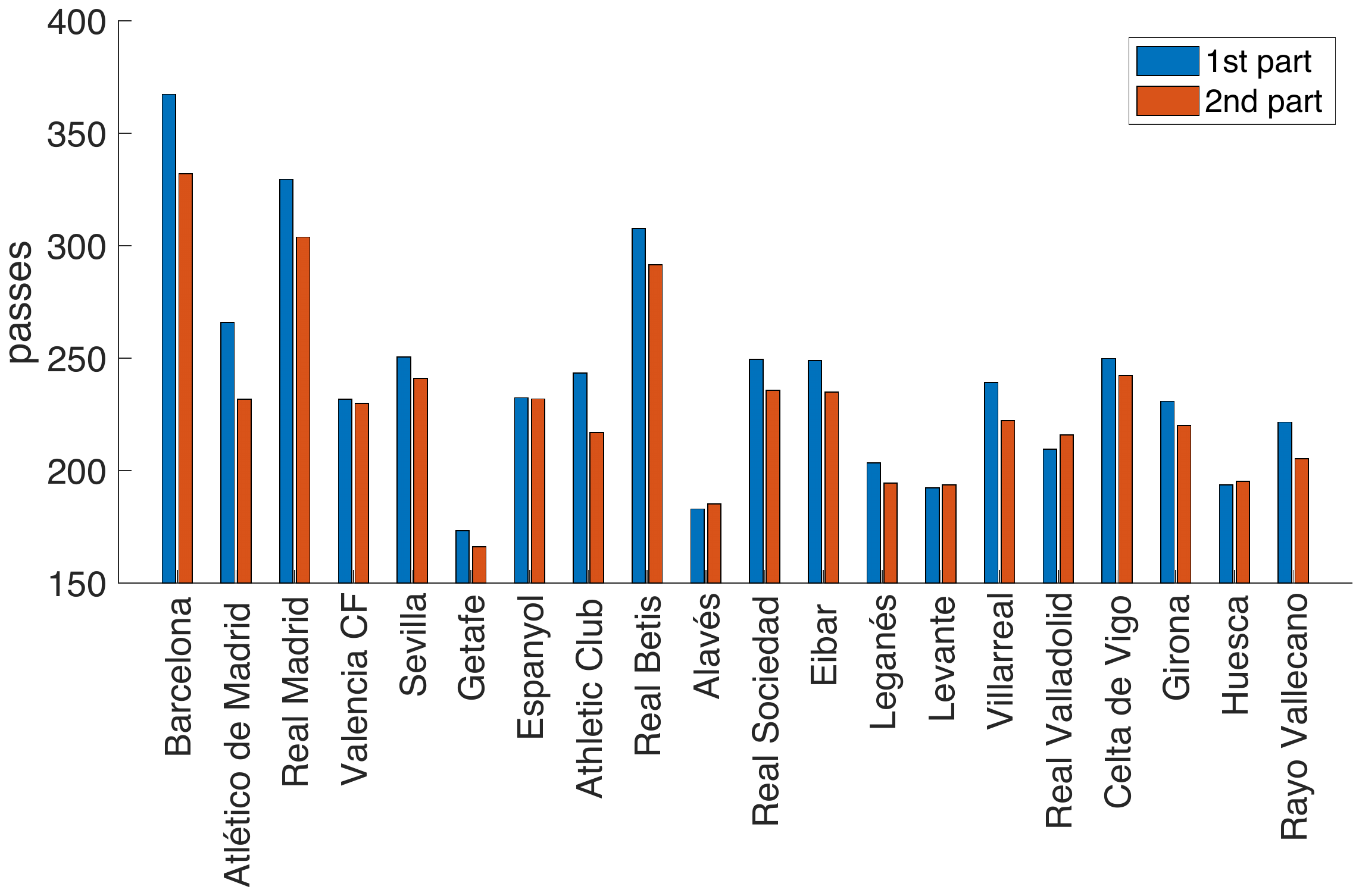}
 \caption{
\textbf{Number of passes per part of the match.} For each team $i$, in blue, number of passes  $n_{1}(i)$ completed during the first part of the match. In red, 
the number of passes  $n_{2}(i)$
completed in the second part. Teams are ordered, from left to right, according to the ranking at the end of the season.
}\label{f02}
\end{figure}

\subsection{Asymmetries between the parts of the match}

Next, we investigated whether the results observed during the whole match were maintained when the two parts of the match were analyzed independently.
In other words, we were interested in finding asymmetries between both halves of a match, in case they exist. 
With this aim, we first analyzed how the number of passes was related to each of the two parts of a match.
In Fig. \ref{f02} we show, for each team $i$, the number of passes at the first and second parts, $n_{1}(i)$ and $n_{2}(i)$, respectively. As we can observe, there is a strong
decrease in the number of passes in the second part of matches. In Fig. \ref{f02}, teams are ordered, from left to right, according to the position at the end of the season. We can observe how 17 teams out of 20 had a lower number of passes in the second part, with Atl\'etico de Madrid and F.C. Barcelona being the teams whose decrease was more pronounced. Only three teams did not display this behavior: Alav\'es, Levante and Huesca.

\begin{figure}[!b] 
 \centering
 \includegraphics[width=0.48\textwidth]{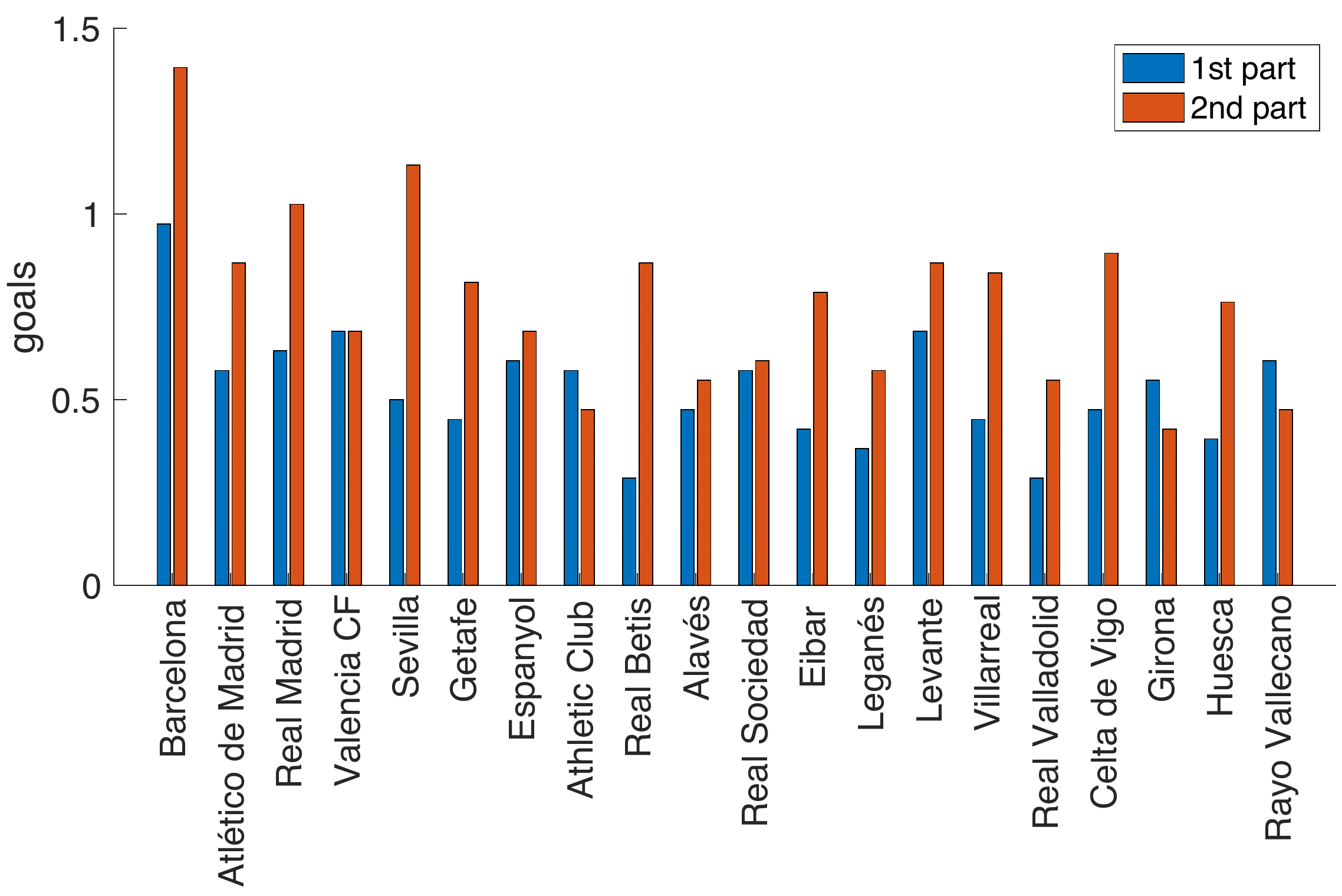}
 \caption{
\textbf{Number of goals per part of the match.} For each team $i$, in blue, the number of goals $m_{1}(i)$ scored during the first part of the match.
 In red, the number of goals $m_{2}(i)$ scored in the second part. Teams are ordered, from left to right, according to the ranking at the end of the season.
 }\label{f03}
\end{figure}

Arriving to this point, a natural question arises: How the reduction of the number of passes is related to the number of goals? 
To answer this question, first, we show in Fig. \ref{f03} the goals scored at each 
part by all teams, i.e. $m_{1}(i)$ and $m_{2}(i)$, respectively. As previously reported in the literature \cite{yiannakos2006,alberti2013,leite2013}, the number of goals increased in the second part. This increase was especially significant for Sevilla and Real Betis, and it is reported at 17 teams. Only Athletic Club, Girona and Rayo Vallecano showed a decrease in the number of goals in the second part. Interestingly, Girona and Rayo Vallecano were relegated at the end of the season. In this way, despite teams completed fewer passes in the second half, they scored more goals, which may seem counterintuitive.

\begin{figure}[!t] 
 \centering
 \includegraphics[width=0.48\textwidth]{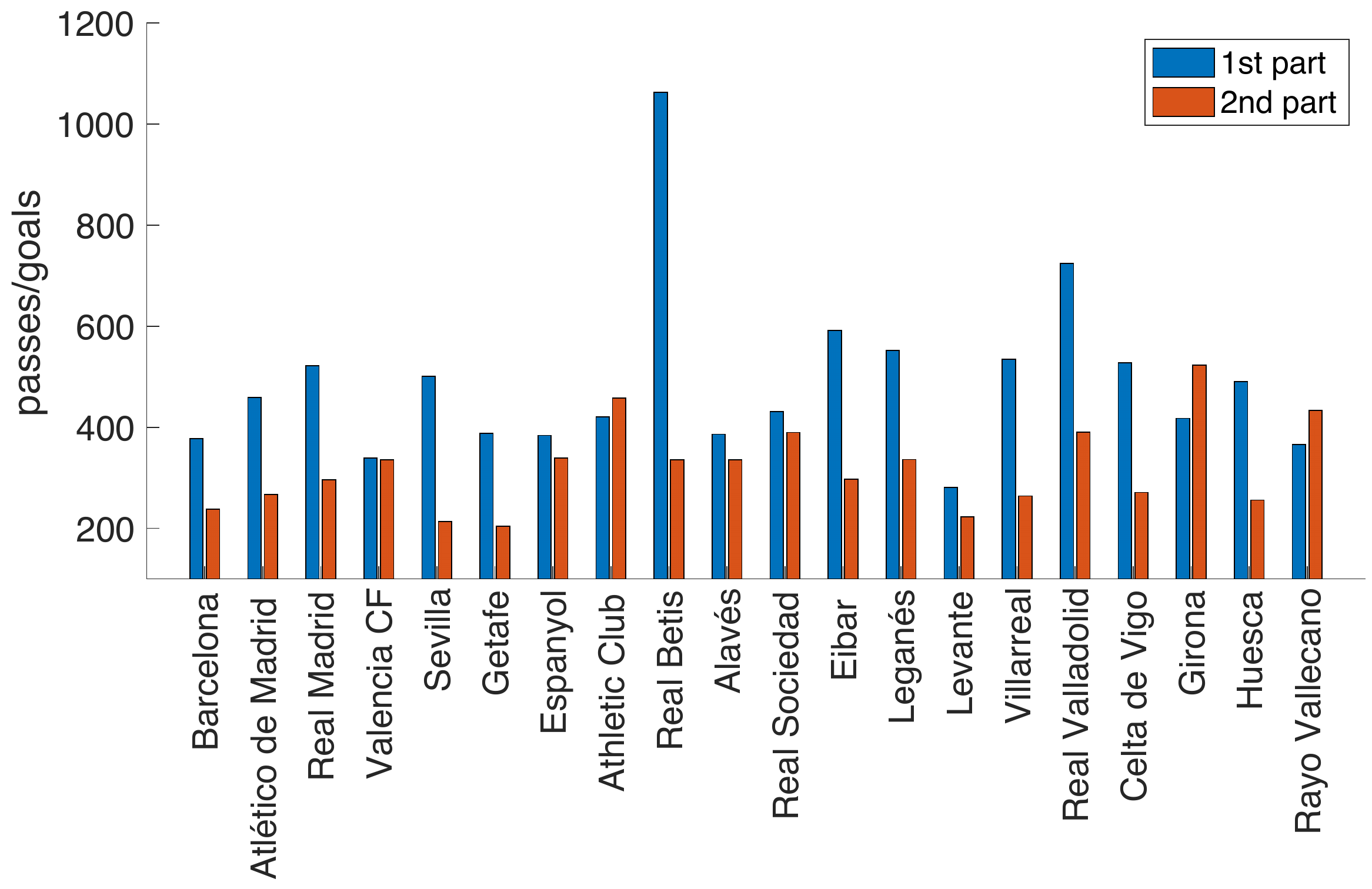}
 \caption{
\textbf{The cost of a goal, in number of passes.} For each team $i$, in blue, the number of passes per scored goal ($n_{1}/m_{1}$) during the first part of the match. In red, the same ratio in the second part  ($n_{2}/m_{2}$). Teams are ordered, from left to right, according to the ranking at the end of the season.
}\label{f04}
\end{figure}

Next, in Fig. \ref{f04}, we divided the total number of passes made at each part by the total number of goals scored by each team. This ratio
is an indicator of how "efficient" passes at each part are or, conversely, how "costly" a pass is in terms of the number of passes. 
Interestingly, we can observe that goals required more passes in the first part of the match for the majority of teams (18 out of 20). Real Betis was the team with the highest differences between parts. The reason is the high number of passes required to score goals in the first parts
of its matches. On the other hand, only two teams deviated from the general behavior: Athletic Club, Girona and Rayo Vallecano.
Finally, it is worth mentioning that Getafe was the team requiring the least number of passes to score a goal. This team has a particular style of play characterized by an intense pressure at higher positions of the field, leading to ball recoveries close to the opponent's goal and, probably, reducing the number of passes before scoring.

\section{Conclusions}
\label{S:3}

Passes and goals are two of the most relevant actions in football. Here, we investigated the interplay between them, showing that there is a strong asymmetry in both the number of passes and goals performed at each part of a match. The analysis of the 20 teams playing at the first division of the Spanish national league showed that there is a moderate correlation between the number of completed passes and the amount of scored goals. When teams were grouped according to their ranking at the end of the season, we observed that the top 4 teams were those making more passes and scoring more goals while, on the contrary, relegated teams had, on average, a lower number of passes and goals. In this way, the first conclusion of our analysis is rather intuitive: Teams making more passes score more goals and, ultimately, 
occupy a higher position at the end of the season.
However, a paradox arises when looking at the distribution of goals between the two parts of a match: While more passes were made during the first half of a match, fewer goals were scored. This fact makes goals more ``costly" in terms of the number of passes during the first part.
The explanation of this paradox is twofold. On the one hand, as discussed in \cite{leite2013}, the decrease in the physical performance of players could be related to a higher probability of making mistakes, which would increase the probability of scoring of any of the two teams. In turn, fatigue could also be responsible for tactic disorganization. On the other hand, the proximity of the end of the game could be a reason for taking more risks in order to change the final result, leading again to an increase in the probability of scoring.

Although we observed that the pass-goal paradox was present at most teams, we must also note that few of them did not fulfill it 
(3 teams out of 20 in our case). Therefore, further
studies should be carried out to investigate (i) why some teams scape from this paradox, (ii) to evaluate its generality by applying a similar
analysis to datasets coming from other football leagues and (iii) to validate the results presented here with larger datasets. 
Finally, other variables, such as playing at home or away, have been shown to influence
the total number of passes and goals \cite{tucker2005} during a match, and they should also be included in the ``to-do" list.

\section*{Acknowledgements}
JMB is supported by MINECO, Spain (FIS2017-84151-P). DRA and DG are funded by Comunidad de Madrid, Spain, through projects MPEJ-2019-AI/TIC-13118 and
PEJ-2018-AI/TIC-11183, respectively.









\end{document}